\documentstyle[twocolumn,twoside]{article}
\newcommand{\bepsilon}{\epsilon \hspace*{-0.8ex} \epsilon}
\newcommand{\blackdee}{\mbox{$\raisebox{-0.15ex}{$\bullet$}
    \hspace*{-1.55ex} \bullet \hspace*{-1.75ex} \partial$}}
\newcommand{\dd}{\raisebox{-0.07ex}{$\bullet$} \hspace{-1.23ex} {\bf d}}
\newcommand{\flux}{\raisebox{-0.2ex}{\Large $\bullet$} \hspace{-2.5ex}\oint}
\newcommand{\jj}{\widetilde{\mbox{\bf \j}}}
\newcommand{\Lagr}{{\cal L \hspace*{-1.45ex} L}}
\title{\Large \bf Five-Dimensional Tangent Vectors in Space-Time \\
\large \bf IV. Generalization of Exterior Calculus}
\author{Alexander Krasulin \\ \it Institute for Nuclear Research of the
Russian Academy of Sciences \\ \it 60th October Anniversary Prospect, 7a,
117312 Moscow, Russia}
\date{\normalsize \bf Abstract \\ \mbox{ } \\ \begin{minipage}{400pt}
\normalsize This part of the series is devoted to the generalization of
exterior differential calculus. I give definition to the integral of a
five-vector form over a limited space-time volume of appropriate dimension;
extend the notion of the exterior derivative to the case of five-vector
forms; and formulate the corresponding analogs of the generalized Stokes
theorem and of the Poincare theorem about closed forms. I then consider
the five-vector generalization of the exterior derivative itself; prove
a statement similar to the Poincare theorem; define the corresponding
five-vector generalization of flux; and derive the analog of the formula
for integration by parts. I illustrate the ideas developed in this paper
by reformulating the Lagrange formalism for classical scalar fields in
terms of five-vector forms. In conclusion, I briefly discuss the five-vector
analog of the Levi-Civita tensor and dual forms. \end{minipage} }
\begin{document}

\maketitle

\begin{flushleft}
A. \it Equivalence classes of two-, three- and \\
\hspace*{2.5ex} four-dimensional volumes
\end{flushleft}
Consider a set $\Re_{2}$ of all smooth parametrized two-dimensional surfaces
going through a fixed space-time point $Q$. Let us lable these surfaces with
calligraphic capital Roman letters: $\cal A, \, B, \, C,$ etc. The two
parameters of surface $\cal A$---its two inner coordinates---will be
denoted as $\lambda^{(1)}_{\cal A}$ and $\lambda^{(2)}_{\cal A}$.

If $f$ is a real scalar function defined in the vicinity of $Q$, one can
evaluate its derivatives at $Q$ relative to the parameters of a given surface
$\cal A$:
\begin{displaymath}
\left[ \frac{d f(P(\lambda^{(1)}_{\cal A},\lambda^{(2)}_{\cal A}))}{d
\lambda^{(k)}_{\cal A}} \right] _{\lambda^{(1)}_{\cal A} =
\lambda^{(1)}_{\cal A}(Q), \; \lambda^{(2)}_{\cal A} =
\lambda^{(2)}_{\cal A}(Q)},
\end{displaymath}
where $k=1\mbox{ or }2$, and let us denote these derivatives as
$\partial^{(k)}_{\cal A} f |_Q$.

Let us focus our attention on the behaviour of two-dimensional surfaces
in the infinitesimal vicinity of $Q$. From that point of view $\Re_{2}$ can
be divided into classes of equivalent surfaces that coincide in direction
or in direction and parametrization. As in the case of parametrized curves,
one can consider several degrees to which two given surfaces, $\cal A$ and
$\cal B$, may coincide:
\begin{enumerate}
   \item The two surfaces have the same direction at $Q$. A more precise
   formulation is the following: there exists a real $2 \times 2$ matrix,
   $\| a^{k}_{\; l} \|$, with a positive determinant and such that for any
   scalar function $f$
\begin{equation}
\partial^{(k)}_{\cal A} f|_{Q} = \sum_{l=1,2} a^{k}_{\; l} \,
\partial^{(l)}_{\cal B} f|_{Q},
\end{equation}
   where $k=1,2$.
   \item The two surfaces have the same direction at $Q$; in the vicinity of
   $Q$ the direction of the corresponding inner coordinate lines is the same;
   and along these lines the corresponding parameters change with equal
   rates. More precisely: for any scalar function $f$
\begin{equation}
\partial^{(k)}_{\cal A} f|_{Q} = \partial^{(k)}_{\cal B} f|_{Q},
\end{equation}
   where $k=1,2$.
   \item In the infinitesimal vicinity of $Q$, the two surfaces have the
   same direction and para\-metrization. This means that
\begin{flushright}
\hspace*{4ex} \hfill $\lambda^{(k)}_{\cal A}(Q) = \lambda^{(k)}_{\cal B}(Q)$
\hfill (3a)
\end{flushright}
   and for any scalar function $f$
\begin{flushright}
\hspace*{4ex} \hfill $\partial^{(k)}_{\cal A} f |_{Q} =
   \partial^{(k)}_{\cal B} f |_{Q},$ \hfill (3b)
\end{flushright} \setcounter{equation}{3}
   where $k=1,2$.
\end{enumerate}

It is a simple matter to check that relations (1), (2) and (3) are all
equivalence relations on $\Re_{2}$, and for each of them one can consider the
corresponding quotient set---the set whose elements are classes of equivalent
two-dimensional surfaces.

It is apparent that one can establish a one-to-one correspondence between
the equivalence classes of parametrized surfaces associated with relation
(2) and ordered pairs of four-vectors, $({\bf A}^{(1)},{\bf A}^{(2)})$,
such that for any surface $\cal A$ from a given class
\begin{displaymath}
\hspace*{10ex} \partial^{(k)}_{\cal A} f|_{Q} = \partial_{{\bf A}^{(k)}}
f|_{Q} \hspace*{5ex} (k = 1,2)
\end{displaymath}
for any scalar function $f$. It is evident that the vectors in a pair
should be linearly independent: otherwise the pair will not correspond to
any nondegenerate surface.

In a similar manner one can establish a one-to-one correspondence between
the classes of equivalent two-surfaces corresponding to relation (3) and
ordered pairs of five-vectors, $({\bf a}^{(1)}, {\bf a}^{(2)})$, according
to the formulae
\begin{displaymath}
\partial^{(k)}_{\cal A} f |_{Q} = \partial_{{\bf a}^{(k)}} f |_{Q} \;
\mbox{ and } \; \lambda^{(k)}_{\cal A} = \lambda_{{\bf a}^{(k)}}
\hspace*{3ex} (k = 1,2)
\end{displaymath}
for any surface $\cal A$ from a given class and for any scalar function
$f$. In this case, besides being linearly independent, the five-vectors
in a pair should satisfy one more requirement: neither of them should be
an element of $\cal E$. Pairs where one of the five-vectors belongs to
$\cal E$ can be considered as a special case corresponding to degenerate
two-dimensional surfaces, which are lines.

\vspace{3ex}

In a similar manner one can consider a set $\Re_{3}$ of all smooth
parametrized three-dimen\-sional hypersurfaces going through a fixed
space-time point $Q$ and a set $\Re_{4}$ of all parametrized four-dimensional
volumes containing $Q$. If $f$ is a real scalar function defined in the
vicinity of $Q$, one can evaluate its derivatives at $Q$ relative to the
parameters $\lambda^{(k)}$ of a given hypersurface or four-volume $\cal A$,
and I will denote these derivatives as $\partial^{(k)}_{\cal A} f |_Q$
(here $k=1,2,3$ for a hypersurface and $k=1,2,3,4$ for a four-volume).

One can then focus one's attention on the behaviour of hypersurfaces and
four-dimen\-sional volumes in the infinitesimal vicinity of $Q$ and consider
different degrees to which two given hypersurfaces or four-volumes may
coincide. The analogs  of relations (2) and (3) will have exactly the
same form, only $k$ will now run 1 through 3 or 1 through 4.

Finally, one can consider the quotient sets corresponding to the latter two
equivalence relations and observe that it is possible to establish a
one-to-one correspondence between their elements and triplets or quadruples
of four- and five-vectors, respectively. The vectors in these triplets and
quadruples should be linearly independent and, in addition, none of the
five-vectors should belong to $\cal E$ unless the corresponding hypersurface
or four-volume is degenerate.

\vspace{3ex} \begin{flushleft}
B. \it Integrals over {\em m}-dimensional volumes
\end{flushleft}
Let us first consider integrals over parametrized curves. Each of them can
be viewed as a rule that assigns a certain number to every finite continuous
parametrized curve within a certain region of space-time. This number is
additive and therefore can be presented as an integral
\begin{equation}
\int^{\lambda_{b}}_{\lambda_{a}} d \lambda \; \phi(\lambda),
\end{equation}
where $\lambda_{a}$ and $\lambda_{b}$ are the end-point values of the curve
parameter and $\phi(\lambda)$ is a certain numerical function, which may
depend on the curve direction in the infinitesimal vicinity of the
integration point.

In applications one is usually interested in invariant integrals, whose value
is independent of the curve parametrization. This is impossible if $\phi$ in
formula (4) depends only on the integration point (if there exists a scalar
function $f$ such that $\phi(\lambda) = f(P(\lambda))$ for every curve),
so the dependence on the curve direction is quite essential. This dependence
can be of different types. Within four-vector exterior calculus one considers
integrals where $\phi(\lambda)$ is a {\em linear} function of the tangent
four-vector ${\bf U} (\lambda)$ and therefore can be presented as a
contraction of ${\bf U} (\lambda)$ with some four-vector 1-form,
$\widetilde{\bf S}$, defined in some region of space-time containing the
curve:
\begin{displaymath}
\phi(\lambda) = \; < \widetilde{\bf S}(P(\lambda)), {\bf U}(\lambda) >.
\end{displaymath}
Such an integral is denoted simply as $\int \widetilde{\bf S}$ and is
referred to as the integral of 1-form $\widetilde{\bf S}$ along the
specified curve.

In a similar way one can consider integrals over limited two-, three- and
four-dimensional volumes. Each of them can be presented in the form
\begin{displaymath}
\int_{\Lambda} d \lambda^{(1)} \ldots d \lambda^{(m)}
\phi(\lambda^{(1)}, \ldots, \lambda^{(m)}),
\end{displaymath}
where $m$ = 2, 3, or 4; $\Lambda$ is the range of variation of inner
coordinates; and $\phi(\lambda^{(1)},\ldots,\lambda^{(m)})$ is a certain
numerical function, which may depend on the direction of the given
$m$-dimensional surface in the infinitesimal vicinity of the integration
point. Within four-vector exterior calculus one deals with integrals whose
integrand is a linear function of each of the tangent four-vectors
${\bf U}^{(k)}$ ($k = 1,\ldots,m$) that correspond to the selected surface
parameters. It is easy to show that for such an integral to be invariant,
the function $\phi$ has to be antisymmetric with respect to permutations of
${\bf U}^{(1)}, \ldots, {\bf U}^{(m)}$. It therefore can be presented as a
contraction of the multivector $\, {\bf U}^{(1)} \wedge \ldots \wedge
{\bf U}^{(m)} \,$ with some four-vector $m$-form $\widetilde{\bf S}$:
\begin{displaymath}
\phi = \; <\widetilde{\bf S} \, , {\bf U}^{(1)} \wedge \ldots \wedge
{\bf U}^{(m)}>.
\end{displaymath}
Such an integral is referred to as the integral of $m$-form
$\widetilde{\bf S}$ over the specified $m$-dimensional surface.

We thus see that in those cases where a given $m$-dimensional surface is
considered only as an integration volume for the integrals discussed above,
it is more adequate to characterize its local direction and parametrization
with the multivector $\, {\bf U}^{(1)} \wedge \ldots \wedge {\bf U}^{(m)} \,$
rather than with the set of $m$ individual tangent four-vectors. One can then
consider the corresponding equivalence relation, according to which two
nondegenerate parameterized $m$-dimensional surfaces going through a
given point $Q$ belong to the same equivalence class if and only if their
multivectors coincide. It is not difficult to see that this relation
is equivalent to relation (1) (and to its analogs for three- and
four-dimensional volumes) with the additional requirement that the matrix
$\| a^{k}_{\; l} \|$ be unimodular. One should also notice that at $m = 1$
this equivalence relation reproduces relation (2) of part II for
parametrized curves.

\vspace{3ex}

Let us now see what will happen if the integration volume is characterized by
tangent {\em five}-vectors. Since we are generalizing exterior calculus, it
is natural to consider the case where the integrand is a linear function of
each of these five-vectors.

Let us, again, start with parametrized curves. Owing to the invariant
decomposition of $V_{5}$ into the direct sum of $\cal Z$ and $\cal E$, any
integral of the considered kind can be presented as
\begin{equation}
\int^{\lambda_{b}}_{\lambda_{a}} d \lambda \; <\widetilde{\bf s},
{\bf u}^{\cal Z}> + \int^{\lambda_{b}}_{\lambda_{a}} d \lambda \;
<\widetilde{\bf s}',{\bf u}^{\cal E}>,
\end{equation}
where ${\bf u}(\lambda)$ is the tangent five-vector and $\widetilde{\bf s}$
and $\widetilde{\bf s}'$ are some five-vector 1-forms. The first term in
this formula is an invariant integral, whose value coincides with that of
the integral $\int \widetilde{\bf S}$ along the same curve, where
$\widetilde{\bf S}$ is the four-vector 1-form that corresponds to
$\widetilde{\bf s}^{\widetilde{\cal Z}}$. The second term in formula (5)
is proportional to
\begin{displaymath}
\int^{\lambda_{b}}_{\lambda_{a}} \lambda \, d \lambda \;
<\widetilde{\bf s}', {\bf 1}>
\end{displaymath}
and is not an invariant integral unless $<\widetilde{\bf s}',{\bf 1}>$ is
identically zero. Therefore, any invariant integral of the considered type
should have the form of the first term in formula (5). The latter will be
referred to as the integral of five-vector 1-form $\widetilde{\bf s}$
along the specified curve.

In a similar way one can deal with invariant integrals over two-, three- and
four-dimen\-sional volumes. One can show that each of them can be presented
in the form
\begin{equation}
\int_{\Lambda} d \lambda^{(1)} \ldots d \lambda^{(m)}
< \widetilde{\bf s},({\bf u}^{(1)})^{\cal Z} \wedge \ldots \wedge
({\bf u}^{(m)})^{\cal Z} >,
\end{equation}
where $m$ = 2, 3, or 4 and $\widetilde{\bf s}$ is now some five-vector
$m$-form. Integral (6) will be denoted as $\int \widetilde{\bf s}$ and
will be referred to as the integral of $m$-form $\widetilde{\bf s}$ over
the specified $m$-dimensional surface. As one can see, only the
$\widetilde{\cal Z}$-component of $\widetilde{\bf s}$ gives contribution
to the integral, so for any integration volume the value of the latter
coincides with that of the integral $\int \widetilde{\bf S}$, where
$\widetilde{\bf S}$ is the four-vector $m$-form that corresponds to
$\widetilde{\bf s}^{\widetilde{\cal Z}}$.

There exists another interesting way of constructing invariant integrals.
For that one should consider the given $m$-dimensional integration surface
(now $m$ = 1, 2, 3 or 4) as a degenerate volume of dimension $m+1$ and
characterize it with the multivector ${\bf u}^{(1)} \wedge \ldots \wedge
{\bf u}^{(m)} \wedge {\bf e}$, where ${\bf u}^{(k)}$ are the five-dimensional
tangent vectors that correspond to nondegenerate inner coordinates and
$\bf e$ is a nonzero five-vector from $\cal E$. The integral itself will
have the form
\begin{equation}
\int_{\Lambda} d \lambda^{(1)} \ldots d \lambda^{(m)} \; < \widetilde{\bf t},
{\bf u}^{(1)} \wedge \ldots \wedge {\bf u}^{(m)} \wedge {\bf e} >,
\end{equation}
where $\widetilde{\bf t}$ is some five-vector form of rank $m+1$.

One should now select the five-vector $\bf e$ from $\cal E$. It is evident
that allowing $\bf e$ to vary from one point to another is equivalent to
introducing a certain weight factor into the integral. As a rule, such
factors are not considered, but even if one does introduce one, it is more
convenient not to absorb it into the multivector, so that the role of the
latter would consist only in specifying the infinitesimal integration volume,
as it does in ordinary exterior calculus. Considering this, it will be taken
that $\bf e$ is a constant vector.

For the same reason one can choose $\bf e$ to have any nonzero size. If
one wishes the five-vector exterior calculus to be applicable to manifolds
without metric and without affine connection (as is its four-vector analog),
one should select the size of $\bf e$ without any reference to the inner
product nor to parallel transport, and the only distinguished choice is
then $\bf e = 1$.

Since ${\bf u}^{(1)} \wedge \ldots \wedge {\bf u}^{(m)} \wedge {\bf 1} =
({\bf u}^{(1)})^{\cal Z} \wedge \ldots \wedge({\bf u}^{(m)})^{\cal Z} \wedge
{\bf 1}$, for any integration volume integral (7) will have the same value
as integral (6) in which $\widetilde{\bf s}$ is defined by the condition
\begin{equation} \left. \begin{array}{l}
<\widetilde{\bf s},({\bf u}^{(1)})^{\cal Z} \wedge \ldots \wedge ({\bf u}^
{(m)})^{\cal Z}> \\ \hspace{10ex} \rule{0ex}{3ex} = \; <\widetilde{\bf t},
{\bf u}^{(1)} \wedge \ldots \wedge {\bf u}^{(m)} \wedge {\bf 1}>.
\end{array} \right. \end{equation}
One should also notice that contrary to the case of integral (6), the
contribution to integral (7) is given by the $\widetilde{\cal E}$-component
of form $\widetilde{\bf t}$.

We thus see that a given five-vector $m$-form $\widetilde{\bf s}$ ($m$ =
1, 2, 3 or 4) can be integrated over volumes of two different dimensions:
$(i)$ over an $m$-dimensional surface, in which case the integral is
determined only by the $\widetilde{\cal Z}$-component of $\widetilde{\bf s}$,
or $(ii)$ over an $(m-1)$-dimensional surface, in which case the integral
depends only on the $\widetilde{\cal E}$-component of $\widetilde{\bf s}$. At
$m=1$ the integration volume in the second case degenerates into an isolated
point (or several isolated points), and the integral is replaced by the value
of the contraction of the given 1-form with the five-vector $\bf 1$.
Five-vector 5-forms can be integrated over volumes of only one
dimension: four. For an obvious reason, they do not have a
$\widetilde{\cal Z}$-component.

  From our analysis it also follows that each additive rule for assigning
numbers to limited $m$-dimensional volumes ($m$ = 1, 2, 3 or 4) produced by
the integrals considered in this section can be interpreted in three
different ways. It can be regarded as an integral of a four-vector
$m$-form; or as an integral of a five-vector $m$-form; or as an integral of
a five-vector $(m+1)$-form, over the considered volume. The three mentioned
forms---which have been denoted above as $\widetilde{\bf S}$,
$\widetilde{\bf s}$, and $\widetilde{\bf t}$, respectively---are related to
one another in the following way:
\begin{displaymath} \left. \begin{array}{l}
< \widetilde{\bf S}, {\bf U}^{(1)} \wedge \ldots \wedge {\bf U}^{(m)} > \\
\hspace{10ex} \rule{0ex}{3ex} = \; < \widetilde{\bf s},
({\bf u}^{(1)})^{\cal Z} \wedge \ldots \wedge ({\bf u}^{(m)})^{\cal Z} > \\
\hspace{10ex} \rule{0ex}{3ex} = \; < \widetilde{\bf t}, {\bf u}^{(1)}
\wedge \ldots \wedge {\bf u}^{(m)} \wedge {\bf 1} >,
\end{array} \right. \end{displaymath}
where ${\bf u}^{(k)} \in {\bf U}^{(k)}$ for $k = 1,\ldots,m$. This invariant
relation between forms of different types exists at each $m$ owing to the
isomorphicity of the three corresponding vector spaces: (1) the space of
multivectors of rank $m$ made out of four-vectors; (2) the space of
multivectors of rank $m$ made out of five-vectors from $\cal Z$; and (3)
the space of wedge products of $\bf 1$ with multivectors of rank $m$ made
out of five-vectors. At $m=1$ these three vector spaces are respectively
$V_{4}$, $\cal Z$, and the maximal vector space of simple bivectors over
$V_{5}$ with the directional vector from $\cal E$, the isomorphisms between
which have already been discussed in part II. In view of the mentioned
isomorphicity, the equivalence relation between parametrized $m$-dimensional
volumes obtained by equating the corresponding multivectors is exactly the
same for all three types of multivectors considered above.

\vspace{3ex} \begin{flushleft}
C. \it Generalized Stokes theorem
\end{flushleft}
Let $\partial V$ be the closed $m$-dimensional boundary of an
$(m+1)$-dimensional surface $V$ and let $\widetilde{\bf S}$ be a four-vector
$m$-form defined throughout $V$. The integral of $\widetilde{\bf S}$ over
$\partial V$, which will be denoted as
\begin{displaymath}
\oint_{\partial V} \widetilde{\bf S},
\end{displaymath}
is called the flux of form $\widetilde{\bf S}$ through the closed surface
$\partial V$. The generalized Stokes theorem states that this flux equals
the integral over the interior of $V$ of a certain $(m+1)$-form, which is
denoted as $\bf d\widetilde{S}$ and is called the exterior derivative of
$\widetilde{\bf S}$:
\begin{equation}
\oint_{\partial V} \widetilde{\bf S} = \int_{V} {\bf d}\widetilde{\bf S}.
\end{equation}

One can give several equivalent definitions to the exterior derivative. As
a first step, one usually defines it for a scalar function (regarded as a
four-vector 0-form): ${\bf d}f$ is such a four-vector 1-form that
\begin{equation}
<{\bf d}f, {\bf U}> \; = \partial_{\bf U} f
\end{equation}
for any four-vector $\bf U$. This enables one to present the basis 1-forms
dual to a coordinate four-vector basis associated with coordinates
$x^{\alpha}$ as exterior derivatives ${\bf d}x^{\alpha}$. One can then
define the effect of $\bf d$ on an arbitrary form in the following way:
for the $m$-form
\begin{equation}
\widetilde{\bf S} = S_{|\alpha_{1} \ldots \alpha_{m}|} \;
{\bf d}x^{\alpha_{1}} \wedge \ldots \wedge {\bf d}x^{\alpha_{m}}
\end{equation}
one has
\begin{equation}
{\bf d\widetilde{S}} = {\bf d}S_{|\alpha_{1} \ldots \alpha_{m}|} \wedge
{\bf d}x^{\alpha_{1}} \wedge \ldots \wedge {\bf d}x^{\alpha_{m}}.
\end{equation}

Another, equivalent definition of the exterior derivative can be given by
induction: for any $m$-form $\widetilde{\bf S}$ and any $n$-form
$\widetilde{\bf T}$
\begin{flushright}
\hspace{5ex} \hfill ${\bf d(\widetilde{S} \wedge \widetilde{T}) =
d\widetilde{S} \wedge \widetilde{T}} + (-1)^{m} {\bf \widetilde{S} \wedge
d\widetilde{T}}$, \hfill (13a)
\end{flushright}
and for any form $\widetilde{\bf S}$
\begin{flushright}
\hspace*{5ex} \hfill $\bf dd\widetilde{S} = 0$. \hfill (13b)
\end{flushright} \setcounter{equation}{13}

One can also define the exterior derivative of any four-vector form
{\em \`{a} la} equation (10): if $\widetilde{\bf S}$ is a 1-form,
then $\bf d\widetilde{S}$ is such a 2-form that
\begin{flushright} \hfill $\left. \begin{array}{l}
\bf <d\widetilde{S},U \wedge V> \; = \; \partial_{U} <\widetilde{S},V>  \\
\bf \hspace*{10ex} - \; \partial_{V}<\widetilde{S},U> - <\widetilde{S},[U,V]>
\end{array} \right.$ \hfill (14a) \end{flushright}
for any four-vector fields $\bf U$ and $\bf V$; if $\widetilde{\bf S}$ is
a 2-form, then $\bf d\widetilde{S}$ is such a 3-form that
\begin{flushright} $\left. \begin{array}{l}
\bf <d\widetilde{S},U \wedge V \wedge W> \\ \bf = \partial_{U} <\widetilde
{S}, V \wedge W> + \; \partial_{V} <\widetilde{S},W \wedge U> \\ \bf + \;
\partial_{W} <\widetilde{S},U \wedge V> - <\widetilde{S}, [U,V] \wedge  W>
\\ \bf - <\widetilde{S},[V,W] \wedge U> - <\widetilde{S},[W,U] \wedge V>
\end{array} \right.$ (14b) \end{flushright} \setcounter{equation}{14}
for any four-vector fields $\bf U$, $\bf V$ and $\bf W$; etc.

Let us now generalize the concept of flux and the Stokes theorem to the
case of five-vector forms.

It is natural to define the flux of a five-vector $m$-form
$\widetilde{\bf t}$ through a closed $m$-dimensional surface $\partial V$
as the integral of $\widetilde{\bf t}$ over $\partial V$, and I will denote
this particular integral as
\begin{displaymath}
\oint_{\partial V} \widetilde{\bf t}.
\end{displaymath}
According to section B, this flux equals the flux through $\partial V$
of the four-vector $m$-form $\widetilde{\bf T}$ corresponding to
$\widetilde{\bf t}^{\widetilde{\cal Z}}$. If $\partial V$ is the boundary
of the $(m+1)$-dimensional volume $V$, then by virtue of the generalized
Stokes theorem for four-vector forms, one has
\begin{displaymath}
\oint_{\partial V} \widetilde{\bf t} = \oint_{\partial V} \widetilde{\bf T}
= \int_{V} {\bf d}\widetilde{\bf T}.
\end{displaymath}
Now, using the one-to-one correspondence that exists between four-vector
forms and $\widetilde{\cal Z}$-components of five-vector forms, one can
present the flux of $\widetilde{\bf t}$ through $\partial V$ as an
integral over the interior of $V$ of a five-vector $(m+1)$-form whose
$\widetilde{\cal Z}$-component corresponds to $\bf d\widetilde{T}$. It is
natural to call this form the exterior derivative of $\widetilde{\bf t}$
and to denote it as $\bf d\widetilde{t}$. We thus obtain the following
variant of the generalized Stokes theorem for five-vector forms:
\begin{equation}
\oint_{\partial V} \widetilde{\bf t} = \int_{V} {\bf d}\widetilde{\bf t},
\end{equation}
where in both integrals the rank of the form equals the dimension of the
integration volume.

It is not difficult to show that if
\begin{equation}
\widetilde{\bf t} = t_{|A_{1} \ldots A_{m}|} \; \widetilde{\bf o}^{A_{1}}
\wedge \ldots \wedge \widetilde{\bf o}^{A_{m}},
\end{equation}
where $\widetilde{\bf o}^{A}$ is the basis of five-vector 1-forms dual to a
{\em passive regular coordinate} five-vector basis, then
\begin{equation}
({\bf d\widetilde{t}})^{\widetilde{\cal Z}} = \partial_{\alpha}
t_{|\alpha_{1} \ldots \alpha_{m}|} \; \widetilde{\bf o}^{\alpha}
\wedge \widetilde{\bf o}^{\alpha_{1}} \wedge \ldots \wedge
\widetilde{\bf o}^{\alpha_{m}}.
\end{equation}
The $\widetilde{\cal E}$-component of $\bf d\widetilde{t}$ is not fixed by
the above correspondence, and to determine it one should consider the flux
of $\widetilde{\bf t}$ through a closed surface of dimension $m-1$. Let us
denote this surface, again, as $\partial V$. The flux in question will then
be the integral of $\widetilde{\bf t}$ over $\partial V$. According to
section B, it equals the flux through $\partial V$ of the five-vector
$(m-1)$-form $\widetilde{\bf s}$ related to $\widetilde{\bf t}$ by an
equation similar to equation (8). If $V$ denotes an $m$-dimensional volume
limited by $\partial V$, then according to the variant of the generalized
Stokes theorem we have obtained above, one has
\begin{displaymath}
\oint_{\partial V} \widetilde{\bf t} = \oint_{\partial V} \widetilde{\bf s}
= \int_{V} {\bf d}\widetilde{\bf s}.
\end{displaymath}
If now one uses the one-to-one correspondence between the
$\widetilde{\cal Z}$-components of five-vector $m$-forms and the
$\widetilde{\cal E}$-components of five-vector $(m+1)$-forms, one can
present the flux of $\widetilde{\bf t}$ through $\partial V$ as an
integral over the interior of $V$ of a five-vector $(m+1)$-form
whose $\widetilde{\cal E}$-component corresponds to $({\bf d}
\widetilde{\bf s})^{\widetilde{\cal Z}}$. This $(m+1)$-form also has the
meaning of an exterior derivative of $\widetilde{\bf t}$, and since its
$\widetilde{\cal Z}$-component is not fixed by the above correspondence,
one can take that this is the same form ${\bf d}\widetilde{\bf t}$ we have
introduced above. We thus obtain another variant of the generalized Stokes
theorem, which has the same form as equation (15), only now in both
integrals the rank of the form is one unit greater than the dimension of
the integration volume. It is not difficult to show that
\begin{displaymath}
({\bf d\widetilde{t}})^{\widetilde{\cal E}} = \partial_{\alpha}t_{|\alpha_{1}
\ldots \alpha_{m-1}|5} \; \widetilde{\bf o}^{\alpha} \wedge \widetilde{\bf
o}^{\alpha_{1}} \wedge \ldots \wedge \widetilde{\bf o}^{\alpha_{m-1}} \wedge
\widetilde{\bf o}^{5},
\end{displaymath}
which together with equation (17) gives one the following general formula
for calculating the exterior derivative of $m$-form (16):
\begin{equation}
{\bf d}\widetilde{\bf t} = \partial_{A} t_{|A_{1} \ldots A_{m}|} \;
\widetilde{\bf o}^{A} \wedge \widetilde{\bf o}^{A_{1}} \wedge \ldots \wedge
\widetilde{\bf o}^{A_{m}}.
\end{equation}

As in four-vector exterior calculus, one can define the exterior derivative
of a five-vector form without any reference to the Stokes theorem, i.e.\ as
a certain operator that produces an $(m+1)$-form out of an $m$-form. One
should, again, start with a scalar function $f$, which is now regarded as a
five-vector 0-form, and define its exterior derivative as such a five-vector
1-form ${\bf d}f$ that
\begin{equation}
<{\bf d}f, {\bf u}> \; = \partial_{\bf u} f
\end{equation}
for any five-vector $\bf u$. One can then present formula (18) in a form
similar to equation (12):
\begin{equation}
{\bf d}\widetilde{\bf t} = {\bf d}t_{|A_{1} \ldots A_{m}|} \wedge
\widetilde{\bf o}^{A_{1}} \wedge \ldots \wedge \widetilde{\bf o}^{A_{m}},
\end{equation}
and take this to be the definition of ${\bf d}\widetilde{\bf t}$.

Another way of defining the exterior derivative for five-vector forms
is similar to equations (14): if $\widetilde{\bf t}$ is a 1-form, then
${\bf d} \widetilde{\bf t}$ is such a 2-form that
\begin{equation} \left. \begin{array}{l}
\bf <d\widetilde{t},u \wedge v> \; = \; \partial_{u} <\widetilde{t},v> \\
\bf \hspace{10ex} - \; \partial_{v}<\widetilde{t},u> - <\widetilde{t},[u,v]>
\end{array} \right. \end{equation}
for any two five-vector fields $\bf u$ and $\bf v$; etc.

Finally, from formula (20) one can derive the analogs of equations (13):
for any $m$-form $\widetilde{\bf s}$ and any $n$-form $\widetilde{\bf t}$
\begin{flushright} \hspace*{5ex} \hfill
${\bf d(\widetilde{s} \wedge \widetilde{t}) = d\widetilde{s} \wedge
\widetilde{t}} + (-1)^{m} {\bf \widetilde{s} \wedge d\widetilde{t}}$,
\hfill (22a) \end{flushright}
and for any form $\widetilde{\bf t}$
\begin{flushright} \hspace*{5ex} \hfill
$\bf dd\widetilde{t} = 0$. \hfill (22b)
\end{flushright} \setcounter{equation}{22}
However, unlike the case of four-vector forms, the latter two equations
(together with equation (19)) are not enough to define the exterior
derivative completely. To gain a better understanding of this fact, one
should recall how things work out in the case of four-vector forms.

  From the generalized Stokes theorem one obtains the following formula for
the exterior derivative:
\begin{equation}
{\bf d\widetilde{S}} = {\bf d}S_{|\alpha_{1} \ldots \alpha_{m}|} \wedge
\widetilde{\bf O}^{\alpha_{1}} \wedge \ldots \wedge \widetilde{\bf
O}^{\alpha_{m}},
\end{equation}
where $\widetilde{\bf O}^{\alpha}$ is any basis of four-vector 1-forms dual
to a coordinate basis. Provided the effect of $\bf d$ on a scalar function
is known, equation (23) is equivalent to equation (13a) and the requirement
\begin{equation}
{\bf d}\widetilde{\bf O}^{\alpha} = 0
\end{equation}
(the latter, by the way, is a necessary and sufficient condition of the
corresponding basis of four-vectors being a coordinate basis). From
equation (10) one can derive that $\widetilde{\bf O}^{\alpha}$ equal
${\bf d}x^{\alpha}$, and so equation (24) follows from equation (13b)
and can be replaced with it.

In the case of five-vector forms, equation (20) is equivalent to equation
(22a) and the requirement
\begin{equation}
{\bf d}\widetilde{\bf o}^{A} = 0,
\end{equation}
which is a necessary condition of the corresponding basis of five-vectors
being a regular coordinate basis. However, from equation (19) one can only
derive that $\widetilde{\bf o}^{\alpha} = {\bf d}x^{\alpha}$. The fifth
basis 1-form, which in this case equals $\jj$, cannot be presented as an
exterior derivative of any 0-form (this is also true of {\em any} basis of
five-vector 1-forms dual to a {\em standard} basis). Thus, its exterior
derivative is not determined by the rule ${\bf dd} = 0$, and to make
equations (22) equivalent to equation (20) one should supplement the
former with a third equation:
\begin{equation}
{\bf d}\jj = 0.
\end{equation}

The fact that basis four-vector 1-forms dual to a coordinate basis can be
presented as exterior derivatives of scalar functions, whereas in the case
of five-vector forms this is possible only for the first four basis 1-forms,
is closely related to the Poincare theorem in application to four-vector
forms and to its analog for forms associated with five-vectors. The Poincare
theorem states that each four-vector $m$-form $\widetilde{\bf S}$ with $m
\geq 1$, which in a certain region of space-time\footnote{This region should
satisfy certain requirements, which can be found e.g.\ in ref.[1].}
satisfies the equation ${\bf d}\widetilde{\bf S} = 0$ (such forms are called
{\em closed}), can be presented in this region as an exterior derivative of
some four-vector $(m-1)$-form. In agreement with this theorem, the basis
four-vector 1-forms $\widetilde{\bf O}^{\alpha}$, which satisfy equation
(24), can be presented as exterior derivatives of certain scalar functions
(which in this particular case can be chosen to coincide with coordinates).
A statement similar to the Poincare theorem can be easily shown to hold for
forms corresponding to five-vectors:
\begin{quote}
Any five-vector $m$-form $\widetilde{\bf s}$ with $m \geq 2$, which in
a certain region of space-time (subject to the same constraints that are
imposed within the Poincare theorem for four-vector forms) satisfies the
equation ${\bf d} \widetilde{\bf s} = 0$, can be presented in this region
as an exterior derivative of some five-vector $(m-1)$-form.
\end{quote}
At $m = 1$ the theorem works for the $\widetilde{\cal Z}$-component only. The
$\widetilde{\cal E}$-component of any five-vector 1-form, if it is nonzero,
is not an exterior derivative of any scalar function; in this case from $\bf
d(\widetilde{s}^{\widetilde{\cal E}}) = 0$ follows $\widetilde{\bf s}^{
\widetilde{\cal E}} = {\rm const} \cdot \jj$. One can easily see how this
theorem manifests itself in the case of basis five-vector 1-forms
$\widetilde{\bf o}^{A}$, which satisfy equation (25).

The representation of basis four-vector 1-forms $\widetilde{\bf O}^{\alpha}$
as exterior derivatives of coordinates is a convenient way of indicating that
the selected basis of 1-forms is dual to a coordinate basis. We see that no
similar convenient representation exists in the case of five-vector forms, so
one should either introduce a special notation for the basis of 1-forms dual
to a regular coordinate basis or each time indicate explicitly what kind of
a basis is used in the formulae presented.

\vspace{3ex} \begin{flushleft}
D. \it Five-vector exterior derivative
\end{flushleft}
In the previous section we have generalized the concept of flux and of
exterior derivative to the case of five-vector forms. It turns out that one
can go farther and consider an operator similar to the exterior derivative
but which differs from the latter in that $\partial$ in the right-hand side
of formula (18) is replaced with its five-vector counterpart: $\partial
+ \lambda \cdot \varsigma \cdot {\bf 1}$. Let us call this operator the
{\em five-vector exterior derivative} and denote its effect on a five-vector
form $\widetilde{\bf t}$ as $\dd \widetilde{\bf t}$.

Since the operator $\partial_{\bf u} + \lambda_{\bf u} \cdot \varsigma \cdot
{\bf 1}$ will often appear in the following formulae, it is convenient to
introduce a special notation for it: $\blackdee_{\bf u}$. By definition, for
any scalar function $f$,
\begin{equation}
\blackdee_{\bf u} f \equiv {\bf u}[f] = \partial_{\bf u} f + \varsigma
\lambda_{\bf u} f.
\end{equation}
  From equations (13) of part II it follows that the effect of
$\blackdee_{\bf u}$ on the product of two scalar functions is formally
described by the rule
\begin{equation}
\blackdee_{\bf u}(fg) = \blackdee_{\bf u}f \cdot g + f \cdot
\blackdee_{\bf u} g - \blackdee_{\bf u}{\it 1} \cdot fg,
\end{equation}
where $\it 1$ is the constant unity function. As in the case of operators
$\partial$ and $\nabla$, it is convenient to introduce the notation
$\blackdee_{A} \equiv \blackdee_{{\bf e}_{A}}$, where ${\bf e}_{A}$ is the
selected basis of five-vectors. Using this notation, one can present the
five-vector exterior derivative of $m$-form (16) as
\begin{equation}
\dd \widetilde{\bf t} = \blackdee_{A} t_{|A_{1} \ldots A_{m}|} \;
\widetilde{\bf o}^{A} \wedge \widetilde{\bf o}^{A_{1}} \wedge \ldots \wedge
\widetilde{\bf o}^{A_{m}},
\end{equation}
where, let us recall, $\widetilde{\bf o}^{A}$ is the basis of five-vector
1-forms dual to a passive regular coordinate basis. One can easily see that
there exists a very simple relation between $\dd$ and $\bf d$: for any form
$\widetilde{\bf t}$
\begin{equation}
\dd \widetilde{\bf t}={\bf d}\widetilde{\bf t} +
\jj \wedge \widetilde{\bf t}.
\end{equation}
  From equation (29) it follows that the effect of $\dd$ on a scalar function
$f$ (regarded as a five-vector 0-form) is given by the formula
\begin{equation}
<\dd f, {\bf u}> \; = \blackdee_{\bf u} f,
\end{equation}
which is the analog of equations (10) and (19) for $\bf d$. One can then
present formula (29) in a form similar to equations (20) and (23):
\begin{equation}
\dd \widetilde{\bf t} = \dd t_{|A_{1} \ldots A_{m}|} \wedge
\widetilde{\bf o}^{A_{1}} \wedge \ldots \wedge \widetilde{\bf o}^{A_{m}},
\end{equation}
which, together with formula (31), can serve as a definition of the effect
of $\dd$ on an arbitrary five-vector form.

Another way of defining the five-vector exterior derivative is similar to
equations (14) and (21): if $\widetilde{\bf t}$ is a five-vector 1-form,
then $\dd \widetilde{\bf t}$ is such a 2-form that
\begin{equation} \left. \begin{array}{l}
\bf <\dd \widetilde{t},u \wedge v> \; = \; \blackdee_{u} <\widetilde{t},v> \\
\bf \hspace{10ex} - \; \blackdee_{v}<\widetilde{t},u> - <\widetilde{t},[u,v]>
\end{array} \right. \end{equation}
for any two five-vector fields $\bf u$ and $\bf v$; etc.

Finally, from formula (32) one can derive the analogs of equations
(13) and (22): for any $m$-form $\widetilde{\bf s}$ and any $n$-form
$\widetilde{\bf t}$
\begin{flushright} \hfill ${\bf \dd (\widetilde{s} \wedge \widetilde{t})
= \dd \widetilde{s} \wedge \widetilde{t}} + (-1)^{m} {\bf \widetilde{s}
\wedge \dd \widetilde{t} - \dd {\it 1} \wedge \widetilde{s} \wedge
\widetilde{t}}$, \hfill (34a) \end{flushright}
and for any form $\widetilde{\bf t}$
\begin{flushright} \hspace*{5ex} \hfill
$\bf \dd \dd \widetilde{t} = 0$. \hfill (34b)
\end{flushright} \setcounter{equation}{34}
These two equations are equivalent to equation (32) and, together with
equation (31), can be used to define the effect of $\dd$ on any five-vector
form by induction. Indeed, by using equation (31) one can show that any
basis of five-vector 1-forms, $\widetilde{\bf o}^{A}$, dual to a passive
regular coordinate basis associated with coordinates $x^{\alpha}$ can be
presented as
\begin{equation} \left \{ \begin{array}{l}
\widetilde{\bf o}^{\alpha} = \dd x^{\alpha} - x^{\alpha} \dd {\it 1} \\
\widetilde{\bf o}^{5} = \dd {\it 1}
\end{array} \right. . \end{equation}
  From equations (34) one then obtains that
\begin{equation}
\dd \widetilde{\bf o}^{A} - \dd {\it 1} \wedge \widetilde{\bf o}^{A}={\bf 0},
\end{equation}
which is nothing but equation (25) expressed in terms of $\dd$. By
induction, from the latter equation one can derive that
\begin{displaymath}
\dd (\widetilde{\bf o}^{A_{1}} \wedge \ldots \wedge
\widetilde{\bf o}^{A_{m}}) - \dd {\it 1} \wedge \widetilde{\bf o}^{A_{1}}
\wedge \ldots \wedge \widetilde{\bf o}^{A_{m}} = {\bf 0}
\end{displaymath}
for any $m \geq 1$, which together with equation (34a) is equivalent to
equation (32).

It is evident that the analog of the Poincare theorem for five-vector forms
presented in the previous section can be reformulated in terms of $\dd$
in the following way: any five-vector $m$-form with $m \geq 2$ or 1-form from
$\widetilde{\cal Z}$, which in a certain region of space-time (subject to
the constraints imposed within the Poincare theorem) satisfies the equation
$\bf \dd \widetilde{s} - \jj \wedge \widetilde{s} = 0$, can be presented
in this region as $\bf \widetilde{s} = \dd \widetilde{t} - \jj \wedge
\widetilde{t}$, where $\widetilde{\bf t}$ is some $(m-1)$-form. There exists
another theorem for $\dd$, which can also be regarded as an analog of the
Poincare theorem:
\begin{quote}
Any five-vector $m$-form $\widetilde{\bf s}$ with $m \geq 1$, which
in a certain region of space-time (subject to the same constraints
that are imposed within the Poincare theorem) satisfies the equation
$\bf \dd \widetilde{s} = 0$, can be presented in this region as a
five-vector exterior derivative of some $(m-1)$-form. At $m = 0$ from
the above equation follows $\bf \widetilde{s} = 0$.
\end{quote}
{\em Proof} : At $m \geq 1$, from equation (30) one has
\begin{equation} \bf
\dd \widetilde{s} = d \widetilde{s} + \jj \wedge \widetilde{s} =
d (\, \widetilde{s}^{\widetilde{\cal Z}}) +
d (\, \widetilde{s}^{\widetilde{\cal E}}) + \jj \wedge
(\, \widetilde{s}^{\widetilde{\cal Z}}) = 0.
\end{equation}
Consequently, $\bf d(\, \widetilde{s}^{\widetilde{\cal Z}}) =
(\dd \widetilde{s})^{\widetilde{\cal Z}} = 0$, and according to the
analog of the Poincare theorem for five-vector forms presented in the
previous section, there exists an $(m-1)$-form $\widetilde{\bf t}$
such that $\bf \widetilde{s}^{\widetilde{\cal Z}} = d
(\, \widetilde{t}^{\widetilde{\cal Z}})$. The $\widetilde{\cal E}$-component
of equation (37) yields $\bf d(\, \widetilde{s}^{\widetilde{\cal E}})
+ \, \jj \wedge (\, \widetilde{s}^{\widetilde{\cal Z}}) =
(\dd \widetilde{s})^{\widetilde{\cal E}} = 0$, so
\begin{equation} \bf
d(\, \widetilde{s}^{\widetilde{\cal E}}) = - \, \jj \wedge \widetilde{s}
^{\widetilde{\cal Z}} = - \, \jj \wedge d(\, \widetilde{t}^{\widetilde{\cal
Z}}) = d(\, \jj \wedge \widetilde{t}^{\widetilde{\cal Z}}).
\end{equation}

At $m \geq 2$, there exists an $(m-1)$-form $\widetilde{\bf r}$
such that $\bf \widetilde{s}^{\widetilde{\cal E}} = \jj
\wedge \widetilde{t}^{\widetilde{\cal Z}} + d
(\, \widetilde{\bf r}^{\widetilde{\cal E}})$, and since $\bf \jj
\wedge \widetilde{r}^{\widetilde{\cal E}} = 0$, one obtains
\begin{displaymath} \bf
\widetilde{s} = \widetilde{s}^{\widetilde{\cal Z}} + \widetilde{s}
^{\widetilde{\cal E}} = d(\, \widetilde{t}^{\widetilde{\cal Z}} +
\widetilde{r}^{\widetilde{\cal E}}) + \jj \wedge (\, \widetilde{t}
^{\widetilde{\cal Z}} + \widetilde{r}^{\widetilde{\cal E}}) = \dd
(\widetilde{t}^{\, \widetilde{\cal Z}} + \widetilde{r}^{\widetilde{\cal E}}).
\end{displaymath}

At $m = 1$, $\widetilde{\bf t}$ is a scalar function, so
$\widetilde{\bf t}^{\widetilde{\cal Z}} = \widetilde{\bf t} \equiv t$. From
equation (38) it then follows that $\widetilde{\bf s}^{\widetilde{\cal E}}
- t \cdot \jj = {\rm const} \cdot \jj$, and since $\bf d({\rm const}) = 0$,
one obtains
\begin{displaymath}
\widetilde{\bf s}=\widetilde{\bf s}^{\widetilde{\cal Z}} + \widetilde{\bf s}
^{\widetilde{\cal E}} = {\bf d}(t + {\rm const}) + (t + {\rm const}) \cdot
\jj = \dd (t + {\rm const}).
\end{displaymath}
Finally, at $m = 0$, one has $\widetilde{\bf s}^{\widetilde{\cal Z}} =
\widetilde{\bf s}$, and the $\widetilde{\cal E}$-component of equation
$\bf \dd \widetilde{s} = 0$ yields $\bf \widetilde{s} = 0 \;$.
\rule{0.8ex}{1.7ex}

It is not difficult to see that the $(m-1)$-form whose five-vector exterior
derivative equals $\widetilde{\bf s}$ is unique only at $m = 1$. At $m \geq
2$ there exists an infinite number of such $(m-1)$-forms, but the difference
of any two of them is a five-vector exterior derivative of some $(m-2)$-form.

\vspace{3ex} \begin{flushleft}
E. \it Five-vector flux and reflected five-vector \\
\hspace*{2ex} exterior derivative
\end{flushleft}
The concept of exterior derivative is directly related to the concept of
flux: by definition, for any five-vector form $\widetilde{\bf s}$, the
exterior derivative $\bf d\widetilde{s}$ is such that its integral over the
interior of any limited volume of appropriate dimension equals the flux of
$\widetilde{\bf s}$ through the volume boundary. We now have a generalization
of $\bf d$: the five-vector exterior derivative $\dd$, and one can use
the above scheme in the opposite direction to obtain the five-vector
generalization of the flux, defining the latter as a quantity which for a
given form $\widetilde{\bf s}$ and any limited volume $V$ of appropriate
dimension, equals the integral of $\dd \widetilde{\bf s}$ over $V$. Let us
call this quantity the {\em five-vector flux of form $\widetilde{\bf s}$
through volume} $V$ and denote it as
\begin{displaymath}
\flux_{V} \widetilde{\bf s}.
\end{displaymath}
The formal analog of the generalized Stokes theorem for $\dd$ can then be
presented as
\begin{equation}
\int_{V} \dd \widetilde{\bf s} = \flux_{V} \widetilde{\bf s}.
\end{equation}

Let us now find the expression for the five-vector flux in the form of an
integral of $\widetilde{\bf s}$. If $\widetilde{\bf s}$ is an $m$-form,
then $\dd \widetilde{\bf s}$ can be integrated over volumes of dimension
$m+1$ and $m$. In the former case the integral will depend only on the
$\widetilde{\cal Z}$-component of $\dd \widetilde{\bf s}$, and since the
latter coincides with $({\bf d} \widetilde{\bf s})^{\widetilde{\cal Z}}$,
the flux corresponding to $\dd$ will be exactly the same as the flux
corresponding to $\bf d$.

The integral of $\dd \widetilde{\bf s}$ over an $m$-dimensional volume $V$
can be presented, by using equation (30) and the generalized Stokes theorem
(15), as
\begin{equation} \begin{array}{l}
\displaystyle \int_{V} \dd \widetilde{\bf s} = \int_{V} {\bf d}
\widetilde{\bf s} + \int_{V} \jj \wedge \widetilde{\bf s} \\
\displaystyle \hspace*{10ex} = \oint_{\partial V} \widetilde{\bf s}
+ (-1)^{m} \int_{V} \widetilde{\bf s} \wedge \jj.
\end{array} \end{equation}
One should now notice that the forms $\widetilde{\bf s}$ and
$\widetilde{\bf t} \equiv \widetilde{\bf s} \wedge \jj$ satisfy equation
(8) of section B, which means that the second integral in the right-hand
side of equation (40) is simply the integral of $\widetilde{\bf s}$ over
$V$. One thus obtains the following expression for the five-vector flux of
$m$-form $\widetilde{\bf s}$ through the $m$-dimensional volume $V$:
\begin{equation}
\flux_{V} \widetilde{\bf s} = \oint_{\partial V} \widetilde{\bf s}
+ (-1)^{m} \int_{V} \widetilde{\bf s}.
\end{equation}
One should notice that unlike the fluxes of five-vector forms considered in
section C, the five-vector flux (41) depends on {\em both} components of
form $\widetilde{\bf s}$.

\vspace{3ex}

An important consequence of the generalized Stokes theorem is the formula
for integration by parts. In the case of five-vector forms, this formula
can be easily obtained by integrating both sides of equation (22a) over a
given volume $V$ and then using equation (15) to convert the integral of
$\bf d(\widetilde{s} \wedge \widetilde{t})$ over $V$ into the integral
of $\bf \widetilde{s} \wedge \widetilde{t}$ over $\partial V$. After
rearranging the terms, one has
\begin{equation}
\int_{V} {\bf d \widetilde{s} \wedge \widetilde{t}} = \oint_{\partial V}
{\bf \widetilde{s} \wedge \widetilde{t}} - (-1)^{m} \int_{V}
{\bf \widetilde{s} \wedge d\widetilde{t}},
\end{equation}
where $m$ is the rank of $\widetilde{\bf s}$. The generalized Stokes theorem
itself can be regarded as a particular case of this formula where
$\widetilde{\bf t}$ is the constant unity 0-form. Let us now derive the
analog of this formula for $\dd$.

Following the same procedure, one integrates both sides of equation (34a)
over a given volume $V$ and then uses equation (39) to convert the integral
of $\bf \dd (\widetilde{s} \wedge \widetilde{t})$ over $V$ into the
five-vector flux of $\bf \widetilde{s} \wedge \widetilde{t}$. After
rearranging the terms, one obtains
\begin{equation} \begin{array}{l}
\displaystyle \int_{V} {\bf \dd \widetilde{s} \wedge \widetilde{t}} \; = \;
\flux_{V} {\bf \widetilde{s} \wedge \widetilde{t}} \\
\displaystyle \hspace*{9ex} - \; (-1)^{m} \! \int_{V} {\bf \widetilde{s}
\wedge \dd \widetilde{t}} + \int_{V} {\bf \, \jj \wedge
\widetilde{s} \wedge \widetilde{t}}.
\end{array} \end{equation}
It is a simple matter to see that for volume $V$ with dimension one unit
greater than the rank of $\widetilde{\bf s} \wedge \widetilde{\bf t}$, this
equation reproduces formula (42). For volume $V$ with dimension equal to
the rank of $\widetilde{\bf s} \wedge \widetilde{\bf t}$, it gives one a new
formula, which, however, is not very useful since the first term in its
right-hand side includes an integral over the interior of $V$ and, in
addition, there exists a third term, which is also an integral over $V$. A
more useful formula can be obtained if in the right-hand side of equation
(43) one isolates the integral over the boundary of $V$ and combines
everything else into a single second term. After simple transformations
one obtains:
\begin{equation} \begin{array}{l}
\displaystyle  \int_{V} {\bf \dd \widetilde{s} \wedge \widetilde{t}} \; = \;
\oint_{\partial V} {\bf \widetilde{s} \wedge \widetilde{t}} \\
\displaystyle \hspace*{9ex} - \; (-1)^{m} \! \int_{V} {\bf \widetilde{s}
\wedge (\dd \widetilde{t}} - 2 \; {\bf \jj \wedge \widetilde{t})}.
\end{array} \end{equation}

We thus see that it makes sense to consider one more operator, which will be
denoted as $\dd^{\star}$ and will be called the {\em reflected five-vector
exterior derivative}. By definition, for any form $\widetilde{\bf t}$,
\begin{equation}
\bf \dd^{\star} \widetilde{t} = d \widetilde{t} - \jj \wedge \widetilde{t}.
\end{equation}
Formula (44) can now be rewritten as
\begin{equation}
\int_{V} {\bf \dd \widetilde{s} \wedge \widetilde{t}} = \oint_{\partial V}
{\bf \widetilde{s} \wedge \widetilde{t}} - (-1)^{m} \int_{V}
{\bf \widetilde{s} \wedge \dd^{\star} \widetilde{t}}.
\end{equation}
It is easy to see that the latter equation can also be presented as
\begin{equation}
\int_{V} {\bf \dd^{\star} \widetilde{s} \wedge \widetilde{t}} =
\oint_{\partial V} {\bf \widetilde{s} \wedge \widetilde{t}} - (-1)^{m}
\int_{V} {\bf \widetilde{s} \wedge \dd \widetilde{t}}.
\end{equation}
Formulae (46) and (47) can be derived directly from the following
equations:
\begin{equation} \left. \begin{array}{rcl}
{\bf d (\widetilde{s} \wedge \widetilde{t})} & = & {\bf \dd \widetilde{s}
\wedge \widetilde{t}} + (-1)^{m} \, {\bf \widetilde{s} \wedge \dd^{\star}
\widetilde{t}} \\ & = & {\bf \dd^{\star} \widetilde{s} \wedge \widetilde{t}}
+ (-1)^{m} \, {\bf \widetilde{s} \wedge \dd \widetilde{t}},
\end{array} \right. \end{equation}
which are analogs of equations (22a) and (34a).

In order to express $\dd^{\star} \widetilde{\bf t}$ in terms of its
components, it is convenient to introduce the corresponding analog
of the operator $\blackdee_{\bf u}$, which I will denote as
$\blackdee^{\star}_{\bf u}$. By definition, for any scalar function $f$
\begin{displaymath}
\blackdee^{\star}_{\bf u} f \equiv \partial_{\bf u} f - \varsigma
\lambda_{\bf u} f.
\end{displaymath}
  From equation (28) one can then derive that
\begin{displaymath}
\partial_{\bf u} (fg) = \blackdee_{\bf u} f \cdot g +
f \cdot \blackdee^{\star}_{\bf u} \, g = \blackdee^{\star}_{\bf u}
f \cdot g + f \cdot \blackdee_{\bf u} \, g,
\end{displaymath}
which is similar to equations (48) for $\bf d$, $\dd$ and $\dd^{\star}$.
Using this new notation, one can present the reflected five-vector exterior
derivative of $m$-form (16) as
\begin{equation}
\dd^{\star} \widetilde{\bf t} = \blackdee^{\star}_{A} t_{|A_{1} \ldots A_{m}
|} \; \widetilde{\bf o}^{A} \wedge \widetilde{\bf o}^{A_{1}} \wedge \ldots
\wedge \widetilde{\bf o}^{A_{m}}.
\end{equation}
One can then derive the analogs of equations (31) through (34), which
will differ from the latter in that everywhere $\dd$ will be replaced
with $\dd^{\star}$ and $\blackdee_{\bf u}$ will be replaced with
$\blackdee^{\star}_{\bf u}$.

\vspace{3ex} \begin{flushleft}
F. \it Euler-Lagrange equations for classical scalar fields
\end{flushleft}
A good illustration to the ideas developed in this paper can be found in
the Lagrange formalism for classical scalar fields. Let us suppose that we
have $N$ such fields and let us denote them as $\phi_{\ell}$ ($\ell$ runs
1 through $N$ and lables the fields, not components). For simplicity let
us confine ourselves to the case where all $\phi_{\ell}$ are real.
Mathematically, the action $S$ corresponding to these fields is an invariant
four-dimensional integral of the type considered in section B. In local
field theory, its integrand depends on the values of the fields and of
their first derivatives at the integration point. Thus, for an arbitrary
four-dimensional volume $V$ one has
\begin{displaymath}
S(V) = \int_{V} \Lagr,
\end{displaymath}
where $\Lagr = \Lagr ({\bf d} \phi_{\ell},\phi_{\ell})$. According to
section B, the Lagrangian density $\Lagr$ can be regarded as a four-vector
4-form, or as a five-vector 4-form, or as a five-vector 5-form. Let us first
recall the traditional interpretation. In this case
\begin{displaymath}
\Lagr = \mbox{$\frac{1}{4!}$} \; {\cal L}_{\alpha \beta \gamma \delta} \;
{\bf d} x^{\alpha} \wedge {\bf d} x^{\beta} \wedge {\bf d} x^{\gamma}
\wedge {\bf d} x^{\delta},
\end{displaymath}
and for a given volume $V$,
\begin{displaymath}
\int_{V} d^{4} \! x \, < \Lagr \, , {\bf E}_{0} \wedge {\bf E}_{1} \wedge
{\bf E}_{2} \wedge {\bf E}_{3} > \; = \int_{V} d^{4} \! x \, {\cal L}_{0123},
\end{displaymath}
where ${\bf E}_{\alpha}$ is the corresponding coordinate four-vector basis.
The equations of motion for the fields $\phi_{\ell}$ are obtained from the
action principle: the physical fields $\phi_{\ell}$ are such that their
variation inside a given volume $V$ with the boundary condition
$\delta \phi_{\ell} |_{\partial V} = 0$ yields a zero first variation of
the action corresponding to $V$. In the standard way, from this principle
one can derive the corresponding Euler-Lagrange equations:
\begin{displaymath}
\partial_{\mu} [ \frac{\partial {\cal L}_{0123}}{\partial (\partial_{\mu}
\phi_{\ell})} ] = \frac{\partial {\cal L}_{0123}}{ \partial \phi_{\ell}},
\end{displaymath}
which, equivalently, can be presented as
\begin{equation} \left. \begin{array}{l}
\frac{1}{4!} \partial_{\mu} \left\{ \frac{\partial {\cal L}_{\alpha \beta
\gamma \delta}}{\partial (\partial_{\mu} \phi_{\ell})} \right\} \, {\bf d}
x^{\alpha} \wedge {\bf d} x^{\beta} \wedge {\bf d} x^{\gamma} \wedge {\bf d}
x^{\delta} \\ \hspace{3ex} \rule{0ex}{4ex} = \frac{1}{4!} \left\{
\frac{\partial {\cal L}_{\alpha \beta \gamma \delta}}{\partial \phi_{\ell}}
\right\} \, {\bf d} x^{\alpha} \wedge {\bf d} x^{\beta} \wedge {\bf d}
x^{\gamma} \wedge {\bf d} x^{\delta}. \end{array} \right. \end{equation}
One should now recall the invariant definition of the derivatives of $\Lagr$
with respect to $\phi_{\ell}$ and ${\bf d} \phi_{\ell}$ and find that
\begin{displaymath}
\left[ \; \frac{\partial \Lagr}{\partial \phi_{\ell}} \; \right] \!
\rule[-1.3ex]{0ex}{2ex}_{\alpha \beta \gamma \delta} = \frac{\partial
{\cal L}_{\alpha \beta \gamma \delta}}{\partial \phi_{\ell}},
\end{displaymath}
so the four-vector 4-form in the right-hand side of equation (50) is exactly
$\partial \Lagr / \partial \phi_{\ell}$. In a similar manner one can find
that the quantities in the curly brackets in the left-hand side of equation
(50) are components of the four-vector-valued 4-form $\partial \Lagr /
\partial ({\bf d} \phi_{\ell})$. One can then use formula (63) of
Appendix to present this side of equation (50) as
\begin{displaymath} \left. \begin{array}{l}
\partial_{\alpha} \left\{ \frac{\partial {\cal L}_{\mu | \beta \gamma \delta
| }}{\partial (\partial_{\mu} \phi_{\ell})} \right\} \, {\bf d} x^{\alpha}
\wedge {\bf d} x^{\beta} \wedge {\bf d} x^{\gamma} \wedge {\bf d} x^{\delta}
\\ \hspace{2ex} \rule{0ex}{4ex} = {\bf d} \left\{ \frac{\partial {\cal L}}
{\partial ({\bf d} \phi_{\ell})} \right\} \! \rule{0ex}{2ex}^{\mu}_{\mu |
\beta \gamma \delta |} \wedge {\bf d} x^{\beta} \wedge{\bf d} x^{\gamma}
\wedge {\bf d} x^{\delta} \; \equiv \; {\bf d} \widetilde{\bf J}^{\ell},
\end{array} \right. \end{displaymath}
where $\widetilde{\bf J}^{\ell}$ is the scalar-valued 3-form obtained from
$\partial \Lagr / \partial ({\bf d} \phi_{\ell})$ by contracting its upper
and its first lower four-vector indices. We thus see that the Euler-Lagrange
equations (50) can be cast into the following abstract form:
\begin{equation}
{\bf d} \widetilde{\bf J}^{\ell} = \widetilde{\bf K}^{\ell},
\end{equation}
where $\widetilde{\bf K}^{\ell} \equiv \partial \Lagr / \partial
\phi_{\ell}$. The integral formulation of this relation is the following:
the flux of the four-vector 3-form $\widetilde{\bf J}^{\ell}$ through the
boundary of any limited four-dimensional volume $V$ equals the integral of
the four-vector 4-form $\widetilde{\bf K}^{\ell}$ over the interior of $V$.

Let us now turn to the other two possible interpretations of $\Lagr$.

In the case where the Lagrangian density is regarded as a five-vector
4-form one obtains practically the same results except that now in all
the formulae all Greek indices are five-vector ones and everywhere $\Lagr$
is replaced with $\Lagr^{\widetilde{\cal Z}}$. Nothing is said about the
$\widetilde{\cal E}$-component of $\Lagr$ and unless some additional ideas
are invokes, it has no relation to the Lagrange formalism.

More interesting results are obtained if $\Lagr$ is regarded as a
five-vector 5-form. In this case
\begin{displaymath}
\Lagr = \mbox{$\frac{1}{5!}$} \; {\cal L}_{ABCDE} \; \widetilde{\bf o}^{A}
\wedge \widetilde{\bf o}^{B} \wedge \widetilde{\bf o}^{C} \wedge
\widetilde{\bf o}^{D} \wedge \widetilde{\bf o}^{E},
\end{displaymath}
and for a given volume $V$,
\begin{displaymath}
\int_{V} d^{4} \! x \, < \Lagr \, , {\bf e}_{0} \wedge {\bf e}_{1} \wedge
{\bf e}_{2} \wedge {\bf e}_{3} \wedge {\bf e}_{5} > \; = \int_{V} d^{4} \!
x \, {\cal L}_{01235},
\end{displaymath}
where ${\bf e}_{A}$ is the corresponding passive regular coordinate basis.
Following the standard procedure, one can obtain the Euler-Lagrange equations
\begin{displaymath}
\partial_{\mu} [ \frac{\partial {\cal L}_{01235}}{\partial (\partial_{\mu}
\phi_{\ell})} ] - \frac{\partial {\cal L}_{01235}}{ \partial \phi_{\ell}}
= 0,
\end{displaymath}
and considering that in a passive regular basis one has $\blackdee_{\mu} =
\partial_{\mu}$ and $\, \blackdee_{5} = 1$, one can rewrite them as
\begin{equation}
\blackdee_{\mu} [ \frac{\partial {\cal L}_{01235}}{\partial (\blackdee_{\mu}
\phi_{\ell})} ] - \blackdee_{5} [ \frac{\partial {\cal L}_{01235}}{\partial
(\blackdee_{5} \phi_{\ell})} ] = 0.
\end{equation}
One should then change the sign of the second term by replacing one of the
$\blackdee_{A}$ with $\blackdee^{\star}_{A}$. This enables one to present
equation (52) as
\begin{displaymath}
\mbox{$\frac{1}{5!}$} \, \blackdee_{H} \left\{ \frac{\partial
{\cal L}_{ABCDE}}{\partial (\blackdee^{\star}_{H} \phi_{\ell})} \right\} \,
\widetilde{\bf o}^{A} \wedge \widetilde{\bf o}^{B} \wedge \widetilde{\bf
o}^{C} \wedge \widetilde{\bf o}^{D} \wedge \widetilde{\bf o}^{E}  = {\bf 0}
\; {\rm (53a)} \end{displaymath}
or as
\begin{displaymath}
\mbox{$\frac{1}{5!}$} \, \blackdee^{\star}_{H} \left\{ \frac
{\partial {\cal L}_{ABCDE}}{\partial (\blackdee_{H} \phi_{\ell})} \right\}
\, \widetilde{\bf o}^{A} \wedge \widetilde{\bf o}^{B} \wedge \widetilde{\bf
o}^{C} \wedge \widetilde{\bf o}^{D} \wedge \widetilde{\bf o}^{E}  = {\bf 0}.
\; {\rm (53b)} \end{displaymath} \setcounter{equation}{53}
One should now recall the invariant definition of the derivatives of
$\Lagr$ and show that the quantities in the curly brackets in equations
(53a) and (53b) are exactly the components of the derivatives
$\partial \Lagr / \partial (\dd^{\star} \phi_{\ell})$ and
$\partial \Lagr /\partial (\dd \phi_{\ell})$, respectively. Finally,
one can use formula (64) of Appendix to present equation (53a) as
\begin{displaymath}
\dd \left\{ \frac{\partial \Lagr}{\partial (\dd^{\star} \phi_{\ell})}
\right\} \! \rule[-1.5ex]{0ex}{4ex}^{H}_{H|ABCD|} \wedge \widetilde{\bf o}
^{A} \wedge \widetilde{\bf o}^{B} \wedge \widetilde{\bf o}^{C} \wedge
\widetilde{\bf o}^{D} = {\bf 0} \; {\rm (54a)} \end{displaymath}
and equation (53b) as
\begin{displaymath}
\dd^{\star} \left\{ \frac{\partial \Lagr}{\partial (\dd \phi_{\ell})}
\right\} \! \rule[-1.5ex]{0ex}{4ex}^{H}_{H|ABCD|} \wedge \widetilde{\bf o}
^{A} \wedge \widetilde{\bf o}^{B} \wedge \widetilde{\bf o}^{C} \wedge
\widetilde{\bf o}^{D} = {\bf 0}, \; {\rm (54b)}
\end{displaymath} \setcounter{equation}{54}
\hspace*{-0.7ex}and it is a simple matter to show that equation (54b)
can be obtained from equation (54a) by changing the sign of the
$\widetilde{\cal Z}$-components of all the quantities involved. We thus
see that the Euler-Lagrange equations in this case can be presented as
\begin{equation}
\dd {\bf \Lambda}^{\ell} = {\bf 0},
\end{equation}
where ${\bf \Lambda}^{\ell}$ is the scalar-valued five-vector 4-form
obtained from $\partial \Lagr / \partial (\dd^{\star} \phi_{\ell})$ by
contracting its upper and its first lower five-vector indices:
\begin{equation}
({\bf \Lambda}^{\ell})_{ABCD} = \left[ \; \partial \Lagr / \partial
(\dd^{\star} \phi_{\ell}) \; \right] \! \rule{0ex}{2ex}^{H}_{HABCD}.
\end{equation}
In the language of integrals, equation (55) means that the five-vector flux
of the 4-form ${\bf \Lambda}^{\ell}$ through any limited four-dimensional
volume $V$ is zero:
\begin{equation}
\flux_{V} {\bf \Lambda}^{\ell} = 0.
\end{equation}

\vspace{3ex} \begin{flushleft}
\rm G. \it Five-vector Levi-Civita tensor and dual forms
\end{flushleft}
As on any other vector space endowed with a nondegenerate inner product,
one can define on $V_{5}$ a completely antisymmetric tensor, $\bepsilon$,
whose rank in this case is five. The magnitude of its only independent
component, say, of $\epsilon_{01235}$, is fixed by the condition:
\begin{center}
$\epsilon_{01235} = +1$ in any orthonormal \\
basis with positive orientation\footnote{The notion of orientation for
five-vector bases is defined in the usual way. In the following I will
consider positive the orientation of a normalized regular basis for which
the associated four-vector basis has positive {\em four}-dimensional
orientation.},
\end{center}
and then the components of $\bepsilon$ in an arbitrary five-vector basis
${\bf e}_{A}$ will be
\begin{displaymath}
\epsilon_{ABCDE} = \eta \cdot |h|^{1/2} \cdot |{\scriptstyle ABCDE}|,
\end{displaymath}
where $h$ denotes the determinant of the matrix $h_{AB} \equiv
h({\bf e}_{A},{\bf e}_{B})$ (at some particular choice of the constant
$\xi$); parameter $\eta$ equals $+1$ if the basis ${\bf e}_{A}$
has positive orientation and $-1$ otherwise; and the symbol
$|{\scriptstyle ABCDE}|$ is defined in the usual way:
\begin{displaymath}
|{\scriptstyle ABCDE}| = \left\{ \begin{array}{rl}
+1, & \mbox{if } ({\scriptstyle ABCDE}) \mbox{ is an even} \\
    & \mbox{permutation of } (01235), \\
-1, & \mbox{if } ({\scriptstyle ABCDE}) \mbox{ is an odd} \\
    & \mbox{permutation of } (01235), \\
 0, & \mbox{ otherwise}.
\end{array} \right. \end{displaymath}
It is easy to show that in any standard basis one has $h = h_{55} \cdot g$,
where $g$ denotes the determinant of the $4 \times 4$ matrix
$g_{\alpha \beta} \equiv g({\bf e}_{\alpha},{\bf e}_{\beta})$,
and therefore in any {\em active} regular basis one has
\begin{displaymath}
\epsilon_{01235} = \eta \cdot |g|^{1/2} \cdot \kappa,
\end{displaymath}
where $\kappa \equiv |\xi|^{1/2}$, and in any {\em passive} regular basis
one has
\begin{displaymath}
\epsilon_{01235} = \eta \cdot |g|^{1/2} \cdot \varpi,
\end{displaymath}
where $\varpi \equiv |\xi|^{1/2} \cdot \varsigma^{-1}$ and $\varsigma$ is
the dimensional constant introduced in section 3 of part II.

As in the case of the four-vector Levi-Civita tensor, it is convenient to
introduce the completely contravariant tensor corresponding to $\bepsilon$,
whose components are
\begin{displaymath}
\epsilon^{ABCDE} = h^{AA'} h^{BB'} h^{CC'} h^{DD'} h^{EE'}
\epsilon_{A'B'C'D'E'},
\end{displaymath}
where matrix $h^{AB}$ is the inverse of $h_{AB}$. It is not difficult to
demonstrate that in any standard basis
\begin{displaymath}
\epsilon^{ABCDE} = - \, {\rm sign} \, \xi \cdot \eta \cdot |h|^{-1/2} \cdot
|{\scriptstyle ABCDE}|,
\end{displaymath}
and therefore
\begin{displaymath}
\epsilon^{ABCDE} \epsilon_{ABCDE} = - \, 5! \cdot {\rm sign} \, \xi \, .
\end{displaymath}
The latter equation is a particular case of the general relation that
expresses the contraction of $\bepsilon$ with its completely contravariant
counterpart over a certain number of indices, in terms of the so-called
permutation tensors:
\begin{displaymath} \left. \begin{array}{l}
\epsilon^{A_{1} \ldots A_{m} C_{1} \ldots C_{5-m}} \epsilon_{B_{1}
\ldots B_{m} C_{1} \ldots C_{5-m}} \\ \hspace{9ex} \rule{0ex}{3ex}
= - \, (5-m)! \cdot {\rm sign} \, \xi \cdot \delta^{A_{1} A_{2} \ldots
A_{m}}_{\hspace*{7.5ex} B_{1} B_{2} \ldots B_{m}},
\end{array} \right. \end{displaymath}
where $m$ can be any integer from 0 to 5 and
\begin{displaymath}
\delta^{A_{1} A_{2} \ldots A_{m}}_{\hspace*{7.5ex} B_{1} B_{2} \ldots B_{m}}
\, \equiv \, m! \; \delta^{[A_{1}}_{\; B_{1}} \delta^{A_{2}}_{B_{2}} \ldots
\delta^{A_{m}]}_{B_{m}}.
\end{displaymath}

Let me say a few words about the differential properties of $\bepsilon$.
Since the normalization of the latter is determined by the inner product
$h$ and since any nondegenerate $h$ is not conserved by parallel transport,
it is {\em a priori} not clear whether or not $\bepsilon$ is a covariantly
constant tensor. One can gain an understanding of the situation from the
following general reasoning.

Let us consider an arbitrary standard orthonormal basis ${\bf e}_{A}$ at
some space-time point $Q$ and let us parallel transport it to a neighbouring
point $Q'$ along some continuous curve $\cal C$ connecting $Q$ with $Q'$.
Let us denote the transported vector ${\bf e}_{A}$ as ${\bf e'}_{A}$.

Owing to the general properties of five-vector parallel transport discussed
in section 3 of part II, ${\bf e'}_{5}$ is, again, a vector from
$\cal E$ and the $\cal Z $-components of ${\bf e'}_{\alpha}$ are orthogonal
one to another and are normalized. Therefore,
\begin{displaymath} \left. \begin{array}{l}
\bepsilon({\bf e'}_{0},{\bf e'}_{1},{\bf e'}_{2},{\bf e'}_{3},{\bf e'}_{5})
\\ \hspace{13ex} \rule{0ex}{3ex} = \; \bepsilon({\bf e'}^{\cal Z}_{0},
{\bf e'}^{\cal Z}_{1},{\bf e'}^{\cal Z}_{2}, {\bf e'}^{\cal Z}_{3},
{\bf e'}_{5}) \\ \hspace{13ex} \rule{0ex}{3ex} = \; (\lambda_{{\bf e'}_{5}} /
\lambda_{{\bf e}_{5}}) \; \bepsilon({\bf e}_{0},{\bf e}_{1},{\bf e}_{2},
{\bf e}_{3},{\bf e}_{5}),
\end{array} \right. \end{displaymath}
and so $\bepsilon$ is covariantly constant or not depending on whether or
not parallel transport conserves the length of the vectors from $\cal E$.
The same result can be obtained by computing the covariant derivative of
$\bepsilon$ in components and finding that in any standard five-vector
basis where the length of the fifth basis vector is constant
\begin{equation}
\epsilon_{ABCDE \, ; \, \mu} = - \, G^{5}_{\; 5 \mu} \epsilon_{ABCDE}.
\end{equation}
Thus, in the case where the connection for five-vectors possesses the local
symmetry described in section 3 of part II, tensor $\bepsilon$ is
covariantly constant.

As its four-vector analog, tensor $\bepsilon$ can be used for converting
multivectors into forms and vice versa. Namely, if $\bf w$ is a multivector
on $V_{5}$ of rank $m$ ($0 \leq m \leq 5$) with components $w^{A_{1} \ldots
A_{m}}$, one can construct from it a five-vector $(5-m)$-form with the
components
\begin{equation}
(m!)^{-1} \, w^{B_{1} \ldots B_{m}} \epsilon_{B_{1} \ldots
B_{m} A_{1} \ldots A_{5-m}} \, .
\end{equation}
To formulate this correspondence in invariant form, one should regard
$\bepsilon$ as an inner product defined for any two multivectors on
$V_{5}$ whose ranks total up to 5, assuming that
\begin{displaymath} \left. \begin{array}{l}
\bepsilon({\bf e}_{A_{1}} \wedge \ldots \wedge {\bf e}_{A_{m}},
{\bf e}_{B_{1}} \wedge \ldots \wedge {\bf e}_{B_{5-m}}) \\ \hspace{14ex}
\rule{0ex}{3ex} \equiv \bepsilon({\bf e}_{A_{1}}, \ldots ,{\bf e}_{A_{m}},
{\bf e}_{B_{1}}, \ldots ,{\bf e}_{B_{5-m}}).
\end{array} \right. \end{displaymath}
Regarding multivectors of rank $m$ as elements of a vector space with
dimension $5! / m! (5-m)!$ and identifying the linear forms on this space
with five-vector $m$-forms, one can employ the same method that has been
used in subsection 3.E of part II to define the maps $\vartheta_{g}$
and $\vartheta_{h}$ and put into correspondence to each multivector $\bf w$
of rank $m$ a certain form $\vartheta_{\epsilon}({\bf w})$ of rank $5-m$
such that
\begin{displaymath} \bf
< \vartheta_{\epsilon}(w) \, , v > \; = \; \bepsilon (w,v)
\end{displaymath}
for any multivector $\bf v$ of rank $5-m$. It is a simple matter to check
that the relation between the components of $\bf w$ and those of the form
$\vartheta_{\epsilon}({\bf w})$ is indeed given by formula (59).

There exists another correspondence between multivectors and forms, which
is determined by the nondegenerate inner product $h$ on $V_{5}$ or, more
precisely, by the inner product of multivectors induced by $h$. In
this case to any multivector $\bf w$ of rank $m$ with components
$w^{A_{1} \ldots A_{m}}$ one puts into correspondence a form of
{\em same} rank with the components
\begin{displaymath}
w^{B_{1} \ldots B_{m}} \, h_{B_{1} A_{1}} \ldots h_{B_{m} A_{m}}.
\end{displaymath}
This latter form, which I will denote as $\vartheta_{h}({\bf w})$, can be
defined invariantly by requiring that for any mulrivector $\bf v$ of rank $m$
\begin{displaymath}
< \vartheta_{h}({\bf w}) \, , {\bf v} > \; = \; h({\bf w,v}),
\end{displaymath}
where $h$ in this case denotes the inner product of multivectors of rank
$m$ induced by the nondegenerate inner product on $V_{5}$.

One can now combine the maps $\vartheta_{\epsilon}$ and $\vartheta_{h}$
and define a one-to-one correspondence between forms or multivectors of
rank $m$ and forms or multivectors of rank $5-m$. The image of an $m$-form
$\widetilde{\bf w}$ with respect to this map will be called a form {\em dual}
to $\widetilde{\bf w}$ and will be denoted as $\widetilde{\bf w}^{\rm dual}$.
It is evident that for any form $\widetilde{\bf w}$
\begin{displaymath}
\widetilde{\bf w}^{\rm dual} \; = \; \vartheta_{\epsilon} \circ
\vartheta^{-1}_{h} (\widetilde{\bf w}),
\end{displaymath}
and the components of $\widetilde{\bf w}^{\rm dual}$ are expressed in terms
of those of $\widetilde{\bf w}$ according to the well-known formula:
\begin{displaymath} \left. \begin{array}{l}
(\widetilde{\bf w}^{\rm dual})_{A_{1} \ldots A_{5-m}} = (m!)^{-1} \,
w_{C_{1} \ldots C_{m}} \\ \hspace{10ex} \rule{0ex}{3ex} \times h^{C_{1} B_{1}}
\ldots h^{C_{m} B_{m}} \, \epsilon_{B_{1} \ldots B_{m} A_{1} \ldots A_{5-m}}.
\end{array} \right. \end{displaymath}
It is easy to prove that
\begin{displaymath}
(\widetilde{\bf w}^{\rm dual})^{\rm dual}
= - \, {\rm sign} \, \xi \cdot \widetilde{\bf w}
\end{displaymath}
and that the duality operation transforms the $\widetilde{\cal Z}$- and
$\widetilde{\cal E}$-components of any form $\widetilde{\bf w}$ respectively
into the $\widetilde{\cal E}$- and $\widetilde{\cal Z}$-components of
$\widetilde{\bf w}^{\rm dual}$. Let me also note that in the case of
five-vector forms, as in the case of forms associated with any other vector
space of odd dimension, the rank of the dual form never equals that of the
initial form, so in the general case one is not able to define the operation
of dual rotation. The only exception are the 2-forms with the zero
$\widetilde{\cal E}$-component, which the duality operation transforms into
3-forms with the zero $\widetilde{\cal Z}$-component. Since there exists
another correspondence between these two types of forms, given by equation
(8) at $m=2$, one is able to define a map similar to the duality operation,
which transforms a 2-form with the zero $\widetilde{\cal E}$-component into
a 2-form of the same type. It is easy to see that such an operation
corresponds to the duality transformation of four-vector 2-forms.

In conclusion, let me mention one useful identity that involves dual forms:
if $\widetilde{\bf s}$ and $\widetilde{\bf t}$ are any two five-vector forms
of same rank, then
\begin{equation}
\widetilde{\bf s} \wedge \widetilde{\bf t}^{\rm dual} \; = \;
\widetilde{\bf s}^{\rm dual} \wedge \widetilde{\bf t} \; = \;
h(\widetilde{\bf s},\widetilde{\bf t}) \cdot \bepsilon,
\end{equation}
where the inner product of forms $\widetilde{\bf s}$ and
$\widetilde{\bf t}$ is defined in the usual way:
\begin{displaymath}
h(\widetilde{\bf s},\widetilde{\bf t}) =
s_{|A_{1} \ldots A_{m}|} \, t^{A_{1} \ldots A_{m}}.
\end{displaymath}

\vspace{3ex} \begin{flushleft}
\bf Acknowledgements
\end{flushleft}
I would like to thank V. D. Laptev for supporting this work. I am grateful
to V. A. Kuzmin for his interest and to V. A. Rubakov for a very helpful
discussion and advice. I am indebted to A. M. Semikhatov of the Lebedev
Physical Institute for a very stimulating and pleasant discussion and to
S. F. Prokushkin of the same institute for consulting me on the Yang-Mills
theories of the de Sitter group. I would also like to thank L. A. Alania,
S. V. Aleshin, and A. A. Irmatov of the Mechanics and Mathematics Department
of the Moscow State University for their help and advice.

\vspace{3ex} \begin{flushleft} \bf
Appendix: Index transposition identity
\end{flushleft}
There exists a very useful identity of purely combinatorial nature, which
enables one to transpose a single index with a group of antisymmetrized
indices if the number of the latter equals the number of values the indices
run through. In the general case, this identity can be formulated as follows:
\begin{quote}
If the array $S_{i \, j_{1} \ldots j_{m}}$ ($m \geq 2$) is completely
antisymmetric in $j_{1}, \ldots ,j_{m}$ and all indices run through the
same $m$ values, then
\end{quote} \begin{equation}
S_{i \, j_{1} \ldots j_{m}} \, = \, m \, (-1)^{m+1} \, S_{[j_{1}
\ldots j_{m}] \, i}.
\end{equation}
{\em Proof} : One has
\begin{equation} \left. \begin{array}{rcl}
S_{i \, j_{1} \ldots j_{m}} & = & S_{i \, [j_{1} \ldots j_{m}]} \\ & = & m
\, S_{[i \, j_{1} \ldots j_{m}]} \; + \; S_{[j_{1} | \, i \, | j_{2}
\ldots j_{m}]} \rule{0ex}{3ex} \\ & & \rule{0ex}{3ex} - \;
\ldots \; + \; (-1)^{m+1} S_{[j_{1} \ldots j_{m}] \, i},
\end{array} \right. \end{equation}
where, as usual, the notation $[j_{1} \ldots j_{k} | \, i \, | j_{k+1}
\ldots j_{m}]$ means antisymmetrization with respect to all the indices
inside the square brackets except for $i$. Since all indices run through
$m$ values only, the first term in the right-hand side of equation (62)
is identically zero. Since $S_{i \, j_{1} \ldots j_{m}}$ is antisymmetric
in its last $m$ indices, one has
\begin{displaymath} \left. \begin{array}{l}
(-1)^{k+1} S_{[j_{1} \ldots j_{k} | \, i \, | j_{k+1} \ldots j_{m}]} \\
\hspace{12ex} \rule{0ex}{3ex} = (-1)^{k+1} (-1)^{m-k} S_{[j_{1} \ldots
j_{m} | \, i \, |]} \\ \hspace{25ex} \rule{0ex}{3ex} = (-1)^{m+1}
S_{[j_{1} \ldots j_{m}] \, i}
\end{array} \right. \end{displaymath}
for all $k$ from 1 to $m-1$, so the remaining $m$ terms in the right-hand
side of equation (62) are all equal, and equation (62) acquires the
form of equation (61). \rule{0.8ex}{1.7ex}

\vspace{3ex}

We will need the following two particular cases of identity (61):
\begin{itemize}
\item If $S^{\mu}_{\alpha \beta \gamma \delta} = S^{\mu}_{[\alpha \beta
\gamma \delta]}$ are components of a four-vector-valued four-vector 4-form,
then
\begin{equation} \left. \begin{array}{l}
\frac{1}{4!} \, \partial_{\mu} S^{\mu}_{\alpha \beta \gamma \delta} =
\frac{1}{3!} \, \partial_{[ \alpha} T_{\beta \gamma \delta ]},
\end{array} \right. \end{equation}
where $T_{\beta \gamma \delta} \equiv S^{\mu}_{\mu \beta \gamma \delta}$.
\item If $S^{H}_{ABCDE} = S^{H}_{[ABCDE]}$ are components of a
five-vector-valued five-vector 5-form, then
\begin{equation} \left. \begin{array}{l}
\frac{1}{5!} \, \blackdee_{H} S^{H}_{ABCDE} = \frac{1}{4!} \, \blackdee_{[A}
T_{BCDE]}.
\end{array} \right. \end{equation}
where $T_{BCDE} \equiv S^{H}_{HBCDE}$.
\end{itemize}

\vspace{6ex} \begin{flushleft}
\bf Reference
\end{flushleft} \begin{enumerate}
 \item L.Schwartz, {\it Analyse Math\'{e}matique}, vol.II, Hermann, 1967.
\end{enumerate}

\end{document}